\documentclass[lettersize,journal]{IEEEtran}
\usepackage{xr}

\makeatletter
\newcommand*{\addFileDependency}[1]{%
\typeout{(#1)}%
\@addtofilelist{#1}
\IfFileExists{#1}{}{\typeout{No file #1.}}
}\makeatother

\newcommand*{\myexternaldocument}[1]{%
\externaldocument{#1}%
\addFileDependency{#1.tex}%
\addFileDependency{#1.aux}%
}

\myexternaldocument{supplement}

\ifCLASSOPTIONcompsoc
  \usepackage[nocompress]{cite}
\else
  \usepackage{cite}
\fi

\ifCLASSINFOpdf
\else
\fi

\usepackage{amsmath,amssymb}
\usepackage{changepage}
\usepackage[flushleft]{threeparttable}
\usepackage{array,booktabs,makecell}
\usepackage{setspace}
\usepackage{algorithm}%
\usepackage{algpseudocode}%
\DeclareMathOperator*{\argmax}{argmax} %
\usepackage{graphicx}

\PassOptionsToPackage{hyphens}{url}\usepackage{hyperref}

\usepackage{relsize}
\usepackage{scalerel}
\usepackage{accents}

\usepackage{bbm}
\usepackage{physics}
\newcolumntype{C}[1]{>{\centering\arraybackslash}p{#1}}
\newcolumntype{L}[1]{>{\raggedright\arraybackslash}p{#1}}

\usepackage{amsthm}
\usepackage{thmtools}
\declaretheoremstyle[
spaceabove=6pt, spacebelow=6pt,
headfont=\normalfont\bfseries,
notefont=\mdseries, notebraces={(}{)},
bodyfont=\normalfont,
postheadspace=0.6em,
headpunct=:
]{mystyle}

\usepackage{caption}
\usepackage{subcaption}

\usepackage{hyperref}
\usepackage{url}
\usepackage{graphicx}
\usepackage{xcolor}

\newcommand{\BOSN}[1]{BOSN#1}
\newcommand*{\rom}[1]{\expandafter\@slowromancap\romannumeral #1@}
\newcommand{\RNum}[1]{\uppercase\expandafter{\romannumeral #1\relax}}
\usepackage{array}
\usepackage{multirow}
\usepackage{amsfonts}
\usepackage{enumerate}
\usepackage{amsthm}
\usepackage{svg}
\usepackage{footnote}
\usepackage{lipsum}       %
\usepackage{changepage}   %

\makesavenoteenv{algorithm}
\counterwithout{footnote}{algorithm}
\makeatletter
\newenvironment{protocol}[1][htb]{%
    \renewcommand{\ALG@name}{Protocol}%
   \begin{algorithm}[#1]%
  }{\end{algorithm}}
\makeatother

\usepackage{tabularx}
\makeatletter
\newcommand{\multiline}[1]{%
  \begin{tabularx}{\dimexpr\linewidth-\ALG@thistlm}[t]{@{}X@{}}
    #1
  \end{tabularx}
}
\makeatother

\makeatletter
\algnewcommand{\LineComment}[1]{\Statex \hskip\ALG@thistlm \(\triangleright\) #1}
\makeatother

\newtheorem{thm}{Theorem}[section]
\newtheorem{defn}[thm]{Definition}

\newtheorem{prop}[thm]{Proposition}

\begin{document}
\title{Mitigating Misinformation Spread on Blockchain Based Online Social Networks}

\author{Rui~Luo, Vikram~Krishnamurthy,~\IEEEmembership{Fellow,~IEEE},
        and~Erik~Blasch,~\IEEEmembership{Fellow,~IEEE}%
\IEEEcompsocitemizethanks{\IEEEcompsocthanksitem R. Luo is with the Sibley School of Mechanical and Aerospace Engineering, Cornell University, Ithaca, NY, 14850. \protect 
E-mail: rl828@cornell.edu
\IEEEcompsocthanksitem V. Krishnamurthy is with the School of Electrical and Computer Engineering, Cornell University, Ithaca, NY, 14850. \protect
E-mail: vikramk@cornell.edu 
\IEEEcompsocthanksitem E. Blasch is with Air Force Office of Scientific Research (AFOSR), Arlington, VA, 22203. \protect
E-mail: erik.blasch.1@us.af.mil
 \IEEEcompsocthanksitem This research was supported by the  U. S. Army Research Office under grants W911NF-19-1-0365, U.S. Air Force Office of Scientific Research under grant FA9550-22-1-0016, and the National Science Foundation under grant CCF-2112457. }}

\maketitle

\begin{abstract}
This paper designs a blockchain protocol to mitigate the spread of misinformation in online social networks. 
The blockchain protocol processes social media postings as transactions, with misinformation being treated as double-spend attacks. 
The probability and duration for a double-spend attack to succeed in the blockchain protocol is used to compute the misinformation propagation time distribution. 
As a result of the proposed protocol, we show that the rate at which misinformation propagates in blockchain based online social networks is inversely correlated to the fraction of honest miners that disapprove the posting. Specifically, we study the dynamics of misinformation propagation in an SIR (Susceptible, Infectious, or Recovered) model with preferential attachment in a multi-community network to account for homophily and community structure in social networks. Numerical experiments using parameters estimated from real-world Twitter hashtag datasets show that the proposed blockchain protocol can lower the number of users exposed by misinformation by delaying the propagation of misinformation.

\end{abstract}


\section{Introduction}
\label{sec:introduction}
The spread of misinformation
puts the information integrity and trust relationships in online social networks (OSNs) at risk. Recent advances in blockchain technology have opened up new possibilities for enabling decentralized trust in a peer-to-peer network \cite{huang2021blockchain}. Blockchain based OSNs (\BOSN{s}) are growing in popularity, with recent examples including Steemit, Sola, Civil, onG.social, and Sapien (Appendix \ref{appendix:list}). In this paper, we examine how blockchain can be used to mitigate the misinformation propagation in OSNs.

An important aspect of \BOSN{s} is that there is no central proprietary authority that stores and controls all the data available to the users. Instead,  data is stored at multiple nodes across  the network.  Since there is no centralized control or moderator of content, such networks are subject to misinformation and inappropriate content. This motivates design of \BOSN{} which uses a blockchain protocol that can mitigate the spread of misinformation.

Blockchain based cryptocurrencies use an immutable, distributed ledger to store the past transactions.
On a similar note, a \BOSN{} can store social media postings on a distributed ledger. The messages are encrypted and recorded in the ledger in an irreversible manner. Additionally, users reach a consensus by approving some messages and rejecting others. %
With these desirable features, blockchain technology has been emphasized as a possible counter-measure to misinformation\cite{note5qayyum2019using, paul2019fake, chen2020incentive}. %

\vspace{0.1in}
\noindent
{\bf Main Results and Organization: }

\noindent(1) Section \ref{sec:model} details our proposed Protocol \ref{alg:BOSN} for \BOSN{s}. 
We model social media postings as transactions and miners detect transactions that constitute a double-spend attack (Definition \ref{defn:double-spend attack}). We develop a conditional probability model for misinformation given a double-spend attack (\ref{eq:cond}). This enables detection of message modifications, yet simply truncating a message does not always imply misinformation. 
Our main result, Theorem \ref{th:misinformation block} in Section \ref{subsec: transmission time}, derives the misinformation propagation time distribution based on the probability and duration of a successful blockchain double-spend attack.

\noindent(2) Section \ref{sec: PA} studies an SIR model with preferential attachment to describe misinformation propagation dynamics in a \BOSN{} with a multi-community structure, which accounts for the Matthew effect and homophily in social networks. Our main result, Theorem \ref{th:preferential attachment}, states that for an SIR model with preferential attachment, misinformation is more likely to spread to a larger network population than an SIR model without preferential attachment.

\noindent(3) Section \ref{sec: numerical} illustrates via numerical studies how misinformation propagates in a \BOSN{} according to an SIR model with preferential attachment. The contact rate and recovery rate of the SIR model are estimated from Twitter hashtag datasets. %
According to simulation results, fewer people are affected by misinformation in \BOSN{s} than in OSNs without blockchain, and the affected users number curve is flattened.

\subsection{Proposed Blockchain Solution to Misinformation}

\noindent(1) {\it Modeling BOSN Messages as Transactions:}  
The proposed BOSN employs a protocol that treats social media postings as transactions. Each transaction involves a source and a message. The source could be a physical location where the news came from, or a URL link to some online resources. %
A \textit{double-spend attack}\cite{brown2021blockchain} occurs when the content of a message is changed, which will be defined formally in Section \ref{subsec: messages}.

A specific group of users called \textit{miners} use their computational resources (e.g., text analysis and data mining) to detect such \textit{double-spend attacks}. 
They determine whether to approve a message. %
The approved message will be included in a block and added to the blockchain. 
The block is confirmed if a certain number of subsequent blocks are added after it\footnote{This confirmation method is similar to the Bitcoin confirmations\cite{nakamoto2008re}, i.e., a number of blocks have been mined in the Bitcoin network since the block that includes the transaction.}. 

\noindent(2) {\it Mitigating the Matthew Effect and Homophily in Misinformation Propagation:}  
In social media, the Matthew effect\cite{west2021misinformationMatthew} occurs when popular users or postings receive disproportionally increasing attention as their fame rises. 
It can be explained by the preferential attachment model\cite{albert2002statisticalBAPA} in which users in a dynamic evolving network form new links preferentially to users with a large number of links. Under the preferential attachment model, misinformation sent from influential users spreads to a larger network population.
In addition, the idea of homophily\cite{mcpherson2001birds} depicts the inclination of people with similar characteristics or interests to connect with each other in OSNs. As a result, misinformation spreads quickly in a group or an echo chamber\cite{luo2021echo}. 

A \BOSN{} is immune to these concerns because the blockchain processes messages in a decentralized manner and message validation is irrespective of the user's popularity or community.

\subsection{Related Work}
We briefly discuss previous works in OSN misinformation in the context of SIR modeling and blockchain (also see listing\footnote{\url{https://www.rand.org/research/projects/truth-decay/fighting-disinformation/search.html}}).

\noindent(1) {\it SIR epidemic model with heterogeneous parameters:}  
The classical SIR model\cite{kermack1932contributions} assumes the same susceptibility for all individuals and the same infectivity for epidemics. %
Previous research has looked at the impact of susceptibility heterogeneity by splitting people into separate groups or constructing a susceptibility distribution. 
Baqaee \cite{note9baqaee2020reopening} considers a five-population SIR model where sub-populations correspond to age groups and the interactions between age groups are calibrated using survey data. 
Gou and Jin \cite{note11gou2017heterogeneous} generalized the SIR model by considering %
heterogeneity of susceptibility and recovery rate. They found that given the mean of the distribution of susceptibility, increasing the variance may block the spread of epidemics.
Lachiany and Louzoun \cite{lachiany2016effects} studied the variability in infection rates which explains the discrepancy between the observation and the predicted number of infected individuals using the classical SIR model.

\begin{table} 
  \caption{Glossary of symbols used in this article}
  \label{table:symbols}
  \centering 
  \resizebox{\linewidth}{!}{
  \begin{threeparttable}
    \begin{tabular}{ C{0.1\textwidth}L{0.5\textwidth} } %
    \toprule
    Symbols  & Description \\ %
    \midrule\midrule
    $G$ & the directed graph representing the OSN \\
    $V$ & the set of nodes \\
    $E$ & the set of edges \\
    $A$ & the adjacency matrix of $G$ \\
    $S(t)$ & the number of susceptible nodes at time $t$ \\
    $I(t)$ & the number of infected nodes at time $t$ \\
    $R(t)$ & the number of recovered nodes at time $t$ \\
    $\mathcal{S}(t)$ & the set of susceptible node and propagation time pairs at time $t$ \\
    $\mathcal{I}(t)$ & the set of infected nodes at time $t$ \\
    $\mathcal{R}(t)$ & the set of recovered nodes at time $t$ \\
    $P$ & the block model matrix of the stochastic block model \\
    $V_m$ & the community $m$, which is a subset of $V$ \\
    $\beta_m$ & the contact rate of nodes in community $m$ \\
    $\alpha_m$ & the recovery rate of nodes in community $m$ \\
    $T_{S\rightarrow I}$ & the propagation time \\
    $P_{T_{S\rightarrow I}}(t)$ & the CDF of $T_{S\rightarrow I}$ \\
    $\dot{P}_{T_{S\rightarrow I}}(t)$ & the PDF of $T_{S\rightarrow I}$ \\
    $b$ & the number of consequent blocks needed to confirm a previous block \\
    $\mu_d$ & the mining rate of dishonest miners \\
    $\mu_h$ & the mining rate of honest miners \\
    $X_i$ & a random variable representing if the $i$-th block is mined by dishonest miners ($X_i=1$) or honest miners ($X_i=-1$) \\
    $p$ & probability that $X_i=1$ \\
    $S_n$ & a random walk representing the mining of the first $n$ blocks \\
    $1-p_0$ & conditional probability that a normal posting constitutes misinformation \\
    $1-p_1$ & conditional probability that a double-spend attack constitutes misinformation \\
    $T_m$ & misinformation propagation time \\ %
    $T_b$ & $S_n$'s first hitting time of $b$ \\
    $T_{-b}$ & $S_n$'s first hitting time of $-b$ \\
    $N_{b}$ & total number of blocks at $T_{b}$ \\ 
    $N_{-b}$ & total number of blocks at $T_{-b}$ \\ %
    $N_d(s)$ & the number of blocks mined by dishonest miners by time $s$ \\
    $N_h(s)$ & the number of blocks mined by honest miners by time $s$ \\
    $N(s)$ & the total number of mined blocks by time $s$ \\
    $N_{i,m}^{(I)}(t)$ & the set of node $i$'s neighbors from community $m$ that are infected at time $t$ \\
    \midrule\midrule
    \end{tabular}
\end{threeparttable}}
\end{table}

\vspace{0.1cm}
\noindent(2) {\it Blockchain based social network models:} 
Blockchain technology has been applied to OSNs to improve data privacy and resilience to misinformation\cite{guidi2020blockchain}. 
Chen et al.\cite{chen2020incentive} proposed a blockchain-based gamification component in social media to incentivize the detection of fake news.
Qayyum et al.\cite{note5qayyum2019using} considered a blockchain implementation of news feed to distinguish facts from fiction. 
Chen\cite{chen2018towards} studied how blockchain can slow down the spread of rumor in OSNs. They explored decentralized contracts and virtual information credits for secure and trustful peer-to-peer information exchange.
Saad et al. \cite{saad2019fighting} proposed a high-level overview of a blockchain enabled framework for misinformation prevention and highlight the various design issues and consideration of such a blockchain enabled framework for tackling misinformation. 

\noindent
Our work differs from existing works in that: 

\noindent (1) We explicitly model misinformation propagation as double-spend attacks in blockchain. With new messages approved by a blockchain transaction approval process, we analytically derive the time distribution that misinformation propagates in \BOSN{}.

\noindent (2) In \BOSN{}, users determine the messages they get and the other users they interact with, rather than depending on service providers (e.g., central authority or official publishers). In other words, users function as blockchain miners for message validation in \BOSN{}, resulting in a decentralized autonomous organization (DAO). 

\noindent (3) When modeling misinformation propagation in \BOSN{s}, we construct a preferential attachment model which accounts for the Matthew effect. %
We show that preferential attachment model causes misinformation to spread to a larger network population.

\noindent (4) In our numerical studies with parameters estimated from real-world Twitter datasets, we show that \BOSN{} slows down the misinformation propagation rate and reduces the number of infected users.

\section{Blockchain Enabled SIR Model for Misinformation Propagation} 
\label{sec:model}
In this section, we describe misinformation propagation in OSNs based on an SIR model in a network. Then, we describe messages as blockchain transactions and misinformation as double-spend attacks. In line with this, a blockchain protocol is developed. 
We derive the misinformation propagation time distribution and incorporate it in the SIR model. Our main idea is to link the propagation time distribution to the time it takes for a double-spend attack to succeed in the blockchain.

\subsection{Propagation Time Distribution in SIR Model}
\label{subsec: decompose} %
In this subsection, we extend the classical Kermack-McKendrick epidemic model\cite{kermack1932contributions} to formulate misinformation propagation dynamics in an OSN. Misinformation propagation is treated as a disease, and users are divided into three states based on how much they are influenced by it: Susceptible ($S$)-the user has not yet received the misinformation, Infected ($I$)-the user is affected by the misinformation and forwards it via its social connections, Recovered ($R$)-the user ignores or recovers from the misinformation. Denote $S(t)$, $I(t), R(t)$ as the number of susceptible, infected, and recovered users at time $t$, respectively. The model has dynamics determined by the following ordinary differential equations:
\begin{equation}
\label{eq:basic SIR}
\begin{split}
    \dot{S}(t) &= -\beta S(t) I(t) \\
    \dot{I}(t) &= \beta S(t) I(t) - \alpha I(t) \\
    \dot{R}(t) &= \alpha I(t)
\end{split}
\end{equation}
with initial conditions $S(0)=S_0, I(0)=I_0, R(0)=R_0$. $N$ denotes the total population size, i.e., $S(t) + I(t) + R(t) = N$. $\beta > 0$ is the contact rate given by mass action incidence\footnote{Mass action incidence is based on the homogeneous mixing assumption. See Section 2.1 in \cite{brauer2019mathematical}.}, i.e., a user is assumed to makes $\beta N$ contacts in unit time. $\alpha > 0$ is the recovery rate at which an infected user recovers. 

The misinformation propagation time, $T_{S\rightarrow I}$, is the time it takes for a contact between a susceptible user and an infected user to result in an infection (See e.g., definition of the generation time in \cite{fraser2007estimating}; sojourn in the exposed stage in \cite{feng2007epidemiological}). In the classical SIR model, $T_{S\rightarrow I}=0$ with probability 1, i.e., users are infected instantly after contact with an infected user.

\begin{figure}
	\centering
	\includegraphics[width=0.45\textwidth]{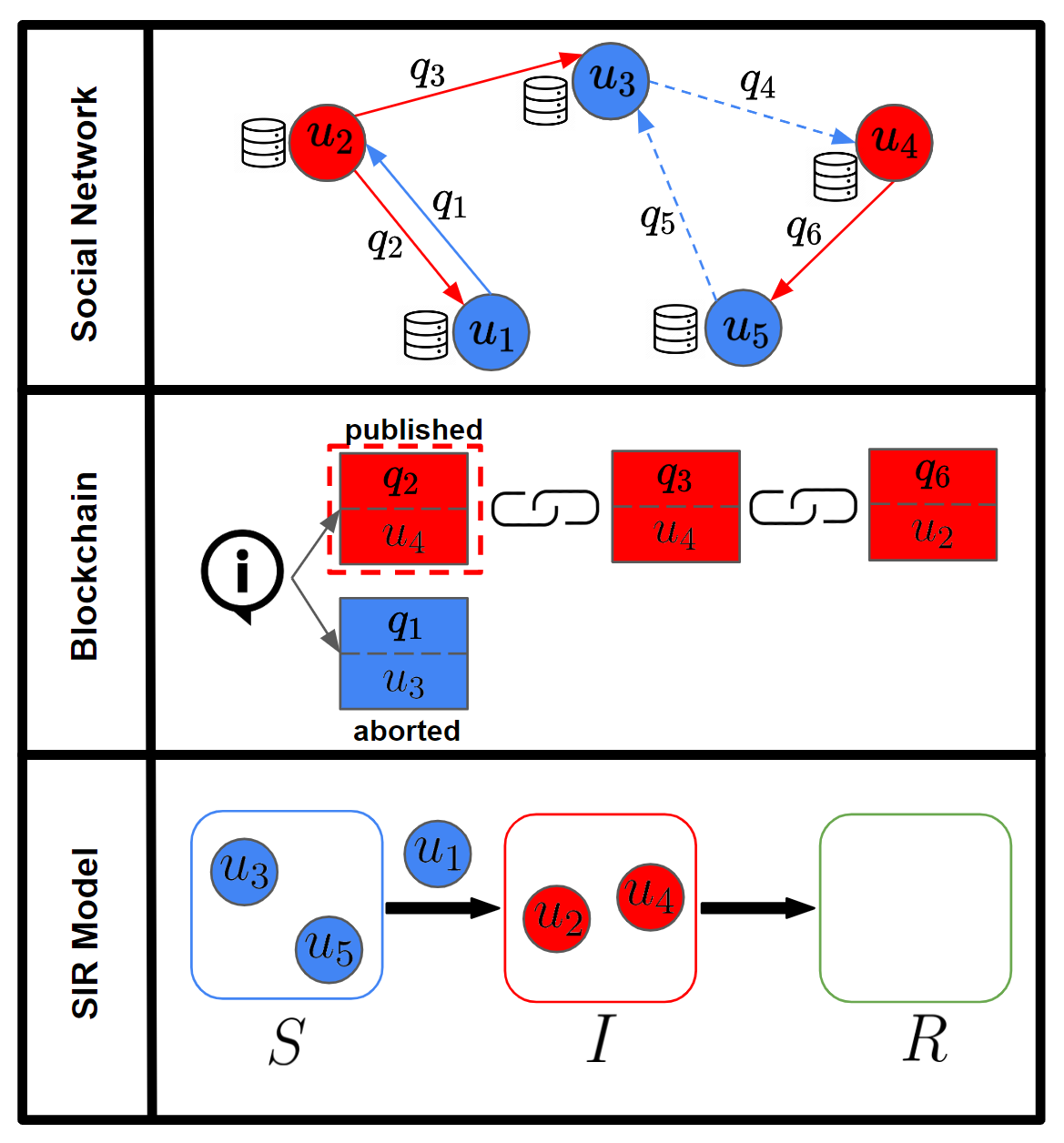}
	\caption{SIR dynamics in a \BOSN{}. \textbf{Social Network:} The top panel shows the \BOSN{}. Red nodes ($u_2,u_4$) and blue nodes ($u_1,u_3,u_5$) represent infected and susceptible users, respectively. Postings are numbered in temporal order ($q_1,\cdots,q_6$) and represented as directed edges. Red ones represent misinformation. Solid edges are approved by miners and added in the blockchain while dashed ones are disapproved. 
	\textbf{Blockchain:} The middle panel illustrates a double-spend attack in blockchain. The original message "i" is turned into truthful message $q_1$ (blue) and misinformation $q_2$ (red), which both get approved (by $u_3$ and $u_4$ respectively as shown in the bottom half of the blocks) and added into blocks. The chain "$q_2 - q_3 - q_5$" outnumbers the other chain "$q_1$" by 2, which leads to $m_2$ published in the \BOSN{}, whereas $m_1$ aborted. \\
    \textbf{SIR Model:} The bottom panel shows the three compartments ($S$-susceptible, $I$-infected, $R$-recovered) in SIR model. Because the misinformation $m_1$ is approved by the blockchain, $u_1$ becomes infected by $u_2$ and enters $I$ from $S$.}
	\label{fig:diagram}
\end{figure}

To account for real-world circumstances in which people become infected after some time rather than instantly, we define $P_{T_{S\rightarrow I}}(t)$ as the cumulative distribution function (CDF) of the propagation time,
\begin{equation} \label{eq:S2I time}
    P_{T_{S\rightarrow I}}(t) = P(T_{S\rightarrow I} \leq t),
\end{equation}
The derivative of (\ref{eq:S2I time}) with respect to time, $\dot{P}_{T_{S\rightarrow I}}(t)$, is the probability density function (PDF) of the propagation time, i.e., the probability that a susceptible user becomes infected after time $t$ since the contact. 

We include the propagation time distribution $P_{T_{S\rightarrow I}}(t)$ in (\ref{eq:basic SIR}), which leads to the following integral equation
\begin{equation}
\label{eq:integral}
    \dot{S}(t) = -\int_{0}^{t} \beta S(\tau) I(\tau) \dot{P}_{T_{S\rightarrow I}}(t-\tau) d\tau
\end{equation}
where $\beta S(\tau) I(\tau) $ is the number of contacts between the susceptible and the infected population at time $\tau$. $\dot{P}_{T_{S\rightarrow I}}(t-\tau)$ is the probability that a susceptible user become infected after $(t-\tau)$ since contact with an infected user.

Note that $\beta$ and $\alpha$ in (\ref{eq:basic SIR}) are both time-invariant. The empirical contact rate and empirical recovery rate, which describe the real-time dynamics of misinformation spread, are defined as follows:
\begin{equation} \label{eq: empirical rate}
    \begin{split}
        \beta(t) & = -\frac{\dot{S}(t)}{S(t) I(t)} \\
        \alpha(t) & = \frac{\dot{R}(t)}{I(t)} \\
    \end{split}
\end{equation}

We further define the misinformation propagation rate as 
\begin{equation} \label{eq: propagation rate}
    \lambda(t) = \frac{\beta(t)}{\alpha(t)} = -\frac{\dot{S}(t)}{S(t) \dot{R}(t)}
\end{equation}
which measures misinformation's rate of effective infections \cite{van2012viral}. %
\vspace{0.2in}

The next sections describe how \BOSN{}'s blockchain protocol leads to a misinformation propagation time and how to determine its distribution analytically.

\subsection{Modeling Messages and Double-Spend Attacks}
\label{subsec: messages}

The \BOSN{} models social media postings as transactions in the blockchain. 
As mentioned in Section 1A, there are two parts to each posting: a source  and a message. While the source identifies the posting's origin, the message is the posting's content. A posting $q$ is represented as a tuple $q = (s, m)$: $s$ is the posting's source, $m$ is the message. A double-spend attack is defined as follows:

\begin{defn}[Double-spend Attack]\label{defn:double-spend attack}
    Double-spend attack is defined as a posting $q_j = (s_j, m_j)$ such that 
    \begin{equation}
    \begin{split}
        & q_j \notin L
        \\
        \textrm{and } & \exists (s_i, m_i) \in L, \textrm{s.t. } s_i = s_{j}, m_i \neq m_{j}    
    \end{split}
    \end{equation}
    where $L=\{q_1, q_2, \cdots\}$ is a blockchain ledger with confirmed postings. In other words, a double-spend attack takes place when a posting has the same source but different content from one that has been confirmed. 
    If $q_j$ is a double-spend attack, we denote $x_{q_j} = 1$; Otherwise, $x_{q_j} = 0$.
\end{defn}

According to Definition \ref{defn:double-spend attack}, a double-spend attack occurs when the news from the same source ($s_j=s_i$) is altered in different contents ($m_j \neq m_i$) to cater to different target audience. 
An example considers the presidential election reported by CNN (assumed to be a neutral news organization) and Drudge Report (a lesser-known, right-leaning news organization). Another example\footnote{\url{https://www.snopes.com/fact-check/university-ban-capital-letters}} is a university memo (same event) from 2018. The misinformation message claims that Leed Trinity University ordered teachers to stop using capital letters, but the truthful message actually states that teachers were instructed to write in a friendly, pleasant tone and avoid “lack of clarity” which could cause anxiety.

Constructing a general model for misinformation is difficult. In this paper we use conditional probabilities to capture the fact that a double-spend attack does not necessarily imply misinformation in terms of conditional probabilities:

\begin{defn}[Conditional Probability Model for Misinformation]\label{defn:conditional probability}
    The conditional probabilities for misinformation are
    \begin{equation}\label{eq:cond}
        p_0 = P(y_q=0|x_q=0), \quad p_1 = P(y_q=0|x_q=1),
    \end{equation}
    where $y_q=1$ indicates that a posting $q$ is misinformation whereas $x_q=1$ indicates that a posting $q$ is double-spend attack (Definition \ref{defn:double-spend attack}).
\end{defn}

Note $p_1 > 0$ ensures that simply truncating a message does not constitute misinformation. Also a double-spend attack does not necessarily imply misinformation. For instance, in the case of an arena's event information, a change to the message's subject might not be a misinformation but rather the announcement of a new event involving a different party.  

\noindent
{\bf Remark.} Although we do not explore the following in this paper, in a real-world \BOSN{}, the conditional probabilities can be estimated as follows:

\begin{enumerate}
    \item \textit{Miner's past track record.} Both the reputation scores and the historical accuracy of miners in disapproving misinformation can be used in determining the $p_0$ and $p_1$. Note that typically $p_0 > p_1$, indicating that double-spend attacks are more likely to constitute misinformation. 
    \item \textit{Word embedding's performance.} $p_0$ and $p_1$ can also be interpreted as the true negative rate and false positive rate of a text-based detector. The detector uses word embedding to transform a message into a two tuple comprising mean vector and covariance matrix of a multivariate Gaussian distribution in Euclidean space, with the covariance matrix serving as a measure of representation uncertainty. The Kullback-Leibler Divergence can be used to evaluate how similar two messages are.
\end{enumerate}

\subsection{Blockchain Protocol}\label{subsec: protocol}
This subsection discusses our proposed blockchain protocol to mitigate the spread of misinformation in a \BOSN{}. A schematic of the \BOSN{} is shown in Fig. \ref{fig:diagram}.

\begin{protocol}
  \caption{Blockchain Protocol for \BOSN{s}}\label{alg:BOSN}
\noindent
{\bf Input: }
A pool of active users from whom to choose to mine the next block.

\noindent
{\bf Output: }
A block addition to the existing blockchain.
  \begin{algorithmic}[1]
    \State A miner is selected uniformly at random from the active users\footnotemark. Either an honest miner (she) or a dishonest miner (he) is selected. 
    \If{the selected miner is honest} \State \multiline{she chooses uniformly at random a posted message that has not been approved and is normal;} 
        \State \multiline{she selects the longest chain of blocks with normal messages.}
    \ElsIf  {the selected miner is dishonest} \State \multiline{he chooses uniformly at random a double-spend attack;}
        \State \multiline{he selects the longest chain of blocks with double-spend attacks.}
    \EndIf
    \State The selected miner adds the new block with the posted message to the end of the selected chain. 
    \If{the selected chain has a $b$-block lead over the other chain} the message in the last $(b+1)$-th block is confirmed. 
    \EndIf
  \end{algorithmic}
\end{protocol}
\footnotetext{The uniformly random selection can extend to a reputation-based election, e.g. \url{https://stackoverflow.com/election}, which is beyond the scope of this paper.}

Each user in a \BOSN{} is also a miner who checks other users' messages before determining whether to add them to the existing blockchain. A \BOSN{} relies on miners to approve communications and disapprove misinformation. 
Miners use their computational resources to query the blockchain ledger $L$ and determine if a message constitutes a double-spend attack. 
The approved message will be included in a block and added to the blockchain. 

We divide miners into two groups: dishonest miners who approve double-spend attacks and honest miners who approve normal messages, i.e., not double-spend attack. Our blockchain protocol is as follows:

According to Protocol \ref{alg:BOSN} and the conditional probability (\ref{eq:cond}), misinformation propagates from an infected user to another user in the following two cases:

\noindent
Case \RNum{1}. Dishonest miners have a $b$-block lead after adding a block containing misinformation with probability $1-p_1$;

\noindent
Case \RNum{2}. Honest miners have a $b$-block lead after adding a block containing misinformation with probability $1-p_0$.

In calculating the time for misinformation to spread, we will account for both scenarios.

\subsection{Propagation Time Distribution with Blockchain}
\label{subsec: transmission time}
In this subsection, we derive the misinformation propagation time distribution in a \BOSN{} using Protocol \ref{alg:BOSN}. 
The distribution accounts for both scenarios given by the conditional probability model (\ref{eq:cond}). 

Denote $\{N_d(t), t\geq 0\}$ and $\{N_h(t), t\geq 0\}$ as the blocks mined by dishonest and honest miners by time $t$, which are independent Poisson processes with respective rates $\mu_d$ and $\mu_h$ \cite{brown2021blockchain}. 

Define i.i.d. random variables $X_i, i=1,2,\cdots$ which take value in $\{+1, -1\}$. The value of $X_i$ indicates whether dishonest miners ($X_i=1$) or honest miners ($X_i=-1$) mined the $i$-th block, and 
\begin{equation}\label{eq:random walk probability}
    p = P(X_i=1) = \frac{\mu_d}{\mu_d + \mu_h}
\end{equation}
A random walk that represents the mining of the first $n$ blocks is defined as
\begin{equation}\label{eq:random walk}
    S_n = \sum_{i=1}^{n} X_i,
\end{equation}

Define $T_b$  as the first time that $N_1$ is $b$ greater than $N_2$, i.e., the number of blocks mined by dishonest miners exceeds that of honest miners by $b$ for the first time,
\begin{equation}
\label{eq:T_k}
    T_b = \textrm{inf}\{t\geq 0: N_d(t)=N_h(t)+b\}, b>0.
\end{equation}
In other words, $T_b$ is the random walk $S_n$'s first hitting time of $+b$. Similarly, denote $T_{-b}$ as $S_n$'s first hitting time of $-b$. 

Define $T_m$ as the misinformation propagation time, 
\begin{equation}\label{eq:T_m}
    T_m = \begin{cases}
        T_b &  \textrm{if $T_b < T_{-b}$ and dishonest miners' initial} \\
        & \textrm{block contains misinformation;} \\
        T_{-b} & \textrm{if $T_b > T_{-b}$ and honest miners' initial} \\
        & \textrm{block contains misinformation;} \\
        \infty & \textrm{otherwise}
    \end{cases}
\end{equation}

\begin{prop}\label{prop:hitting probability}
    The probability for the (biased) random walk $S_n$ (\ref{eq:random walk}) to hit $+b$ before $-b$ is
    \begin{equation}\label{eq:Tb<T-b}
    \begin{split}
        P(T_b < T_{-b}) = \frac{1 - (\frac{\mu_h}{\mu_d})^b}{1 - (\frac{\mu_h}{\mu_d})^{2b}} = \frac{1}{1 + (\frac{\mu_h}{\mu_d})^b}
    \end{split}
    \end{equation}
\end{prop}
\begin{proof}
    See Appendix \ref{appendix: proposition 1}.
\end{proof}

We now derive the probability distribution of the misinformation propagation time in the two aforementioned scenarios. 

\vspace{0.2in}
\noindent
{\bf Case \RNum{1}. %
\emph{Dishonest miners have a $b$-block lead after adding a block containing misinformation with probability $1-p_1$.}}

In case \RNum{1}, $T_b < T_{-b}$. Let \textit{misinformation propagation block} be  
\begin{equation}\label{eq:N_b}
    N_b = \textrm{min}\{n:S_n = b \wedge (S_i > -b \ \forall 1\leq i<n) \},
\end{equation}
which represents the total number of mined blocks when misinformation propagates in case \RNum{1}.

We show the probability mass function (PMF) of $N_b$ in the following Theorem \ref{th:misinformation block}.

\begin{thm}[PMF of the misinformation propagation block in case \RNum{1}]\label{th:misinformation block}
    In case \RNum{1} of misinformation propagation, the misinformation propagation block is equivalent to the random walk $S_n$'s first hitting time of $+b$, and its PMF is: 
    \begin{equation}\label{eq:first hitting case 1}
    \begin{split}
        &P(N_b=b+2i)=  p^{b+i} (1-p)^{i} \biggl[\binom{b+2i-1}{i} - \binom{b+2i-1}{i-1} \\
        &  - \binom{b+2i-1}{i-b} + \binom{b+2i-1}{i-b-1} + \binom{b+2i-1}{i-2b} \biggr] , i \in \mathbb{N},
    \end{split}
    \end{equation}
    where $p$ is defined in (\ref{eq:random walk probability}).
\end{thm}
\begin{proof}
    See Appendix \ref{appendix: theorem 1}.
\end{proof}

\noindent
{\bf Case \RNum{2}. %
\emph{Honest miners have a $b$-block lead after adding a block containing misinformation with probability $1-p_0$.}}

Similar to case \RNum{1}, let 
\begin{equation}
    N_{-b} = \textrm{min}\{n:S_n = -b \wedge (S_i < b \ \forall 1\leq i<n) \},
\end{equation}

PMF of $N_{-b}$ is obtained by switching $p$ and $1-p$ in (\ref{eq:first hitting case 1}),
\begin{equation}\label{eq:first hitting case 2}
\begin{split}
    &P(N_{-b}=b+2i)=  (1-p)^{b+i} p^{i} \biggl[\binom{b+2i-1}{i} - \binom{b+2i-1}{i-1} \\
    &  - \binom{b+2i-1}{i-b} + \binom{b+2i-1}{i-b-1} + \binom{b+2i-1}{i-2b} \biggr] , i \in \mathbb{N}
\end{split}
\end{equation}

Denote the PMFs in both scenarios (\ref{eq:first hitting case 1}, \ref{eq:first hitting case 2}) as
\begin{equation}\label{eq:gb and g_b}
\begin{split}
    g_{b}(i) &=P(N_b=b+2i), \\
    g_{-b}(i) &=P(N_{-b}=b+2i), \\
\end{split}
\end{equation}
Denote the hitting probability of the random walk (\ref{eq:Tb<T-b}) as 
\begin{equation}\label{eq:hitting probability}
    p_{b} =P(T_b < T_{-b}), \quad p_{-b} =P(T_b > T_{-b})=1-p_{b}.
\end{equation}

The following Theorem gives an expression for the PMF of the misinformation propagation time $T_m$ (\ref{eq:T_m}).

\begin{thm}
The PMF of misinformation propagation time, denoted as $P(T_m=t)$, is the sum of the probabilities of the joint events that (1) a total of $(j+b)$ blocks being mined; and (2) miners mining the $(j+b)$-th block, resulting in their chain outnumbering the other miners' chain for the first time by $b$ blocks, where $j\in \mathbb{N} $:
\begin{equation}
\label{eq:time distribution blockchain}
\begin{split}
    &P(T_m = t) \\ 
    =& \sum_{j=0}^{\infty} \bigl[ (1-p_1) p_b g_b([j/2]) + (1-p_0) p_{-b} g_{-b}([j/2]) \bigr]  \frac{e^{-\mu t} (\mu t)^{j+b}}{(j+b)!}  
\end{split}
\end{equation}
where $p_0$ and $p_1$ are the conditional probabilities defined in (\ref{eq:cond}), $p_b$ and $p_{-b}$ are the hitting probabilities defined in (\ref{eq:hitting probability}), $g_b$ and $g_{-b}$ are defined in (\ref{eq:gb and g_b}), $\mu = \mu_d + \mu_h$, and $[x]$ denotes the largest integer less than or equal to $x$.
\end{thm}
\begin{proof}
    See Proof of Proposition 1 in \cite{brown2021blockchain}.
\end{proof}

The key takeaway from Theorem \ref{th:preferential attachment} is that misinformation takes longer to spread than truthful message and the misinformation propagation time distribution depends on the proportion of dishonest miners $\frac{\mu_h}{\mu_h + \mu_d}$, the number of confirmation blocks $b$, and the conditional probability $p_0, p_1$,  as described in Protocol \ref{alg:BOSN} and (\ref{eq:cond}).

\subsection{Difference Equations for SIR Model}
\label{subsec:discrete}
In this subsection, we time discretize the model in  (\ref{eq:integral}) (Section \ref{subsec: decompose}) which yields difference equations for computer simulation in Section \ref{subsec:numerical}. 

First, we relate the PMF  of $T_m$  (\ref{eq:time distribution blockchain}) to the propagation time distribution $P_{T_{S\rightarrow I}}$ in (\ref{eq:S2I time})
\begin{equation}
\label{eq:discrete time distribution}
    \begin{split}
        P_{T_{S\rightarrow I}}(t) &= P(T_m \leq t) \\
        \dot{P}_{T_{S\rightarrow I}}(t) &= P(T_m = t) 
    \end{split}
\end{equation}

The integral equation (\ref{eq:integral}) can be expressed as
\begin{equation}
\label{eq:integral 3}
    \begin{split}
        \dot{S}(t) &= - \int_{0}^{t} \beta S(\tau) I(\tau) P(T_m = t-\tau) d\tau \\
        &= - \int_{0}^{t} \beta S(\tau) I(\tau) \sum_{j=0}^{\infty}
         \bigl[ (1-p_1) p_b g_b([j/2]) \\ 
        &\phantom{=}  + (1-p_0) p_{-b} g_{-b}([j/2]) \bigr]  \frac{e^{-\mu (t-\tau)} (\mu (t-\tau))^{j+b}}{(j+b)!}  d \tau \\
    \end{split}
\end{equation}
where $[x]$ denotes the largest integer less than or equal to $x$. %

We time discretize (\ref{eq:integral 3}) resulting in the difference equation
\begin{equation} \label{eq: discrete 1}
     S(t+1) = S(t) - \sum_{i=0}^{t} \beta S(i) I(i)  P\Big(t-i \leq T_m < t-i+1 \Big) 
\end{equation}

Similarly, the infected and recovered population evolve according to the difference equations
\begin{equation} \label{eq: discrete 2}
    I(t+1) = I(t) + \sum_{i=0}^{t} \beta S(i) I(i) P\Big(t-i \leq T_m < t-i+1 \Big) - \alpha I(t) 
\end{equation}
\begin{equation}
\label{eq:recovery rate}
    R(t+1) = R(t) + \alpha I(t)
\end{equation}

Note that $\mu$ in (\ref{eq:integral 3}) is the combined mining rate of all dishonest and honest miners and indicates the speed of block mining. We assume that block mining is faster than message propagation in order to provide \BOSN{} users with a feeling of system reacting instantaneously and swiftly. Indeed, the blockchain technology is capable to meet this requirement, with Stellar and Solana achieving 2 to 4 seconds per block\footnote{\url{https://alephzero.org/blog/what-is-the-fastest-blockchain-and\%2Dwhy-analysis-of-43-blockchains/}}. On the other hand, message propagation on Twitter or Facebook may take minutes.

\section{Preferential Attachment Model and Multi-Community Network}
\label{sec: PA}
In this section, we consider a \BOSN{} which has a multi-community network structure. We also construct a preferential attachment model to account for popular users' disproportionate role in misinformation propagation. 
Recall the definition of misinformation in Definition 2. %
The aim of this section is to show how social network topology affects the misinformation propagation rate and the number of users infected by misinformation (\ref{eq: discrete 1}). The next section demonstrates how well the proposed blockchain protocol works to reduce the misinformation propagation in multi-community networks.

\subsection{Misinformation Propagation in a Multi-community Network}
\label{subsec: multi community without}
To account for the homophily and community structure of social networks, we generalize the SIR dynamics (\ref{eq:basic SIR}) from a homogeneous population to a multi-community network. We employ a stochastic block model, with users divided into communities with varying contact rates and recovery rates. Then, we derive the corresponding state transition of the SIR model. 

Consider a network represented by a directed graph $G=(V,E)$, where $V$ is the node set representing the users and $E$ is the edge set representing the users' friendship.  The adjacency matrix $A=[A_{ij}]_{N\times N}$ is a binary valued matrix where
\begin{equation}
    A_{ij} = 
    \begin{cases}
        1, & \textrm{there is an edge from $i$ to $j$} \\
        0, & \textrm{otherwise} \\
    \end{cases}
\end{equation}

In the multi-community model, $V$ is partitioned into $M$ communities where each node belongs to one community, i.e., $V=V_1 \cup \cdots \cup V_M$ and $V_i \cap V_j = \emptyset, \forall i, j \in \{1, \cdots, M\}, i\neq j$. 
Just as population in different age groups have different infection fatality rate to the epidemic\cite{acemoglu2021optimal}, users in different communities have different contact rates and recovery rates to misinformation, i.e., contact rates $\beta_1 < \cdots < \beta_M$ and recovery rates $\alpha_1 > \cdots > \alpha_M$, where $\beta_m$ and $\alpha_m$ denote the rates for users in community $V_m, m\in \{1, \cdots, M\}$.  

Let $c_i$ denotes node $i$'s community label, e.g. $c_i=m$ if $i \in V_m$. We also define the block model matrix $P$ as a symmetric $M\times M$ matrix which specifies the probability that two nodes connect based on their communities, i.e.,
\begin{equation}
\label{eq:SBM}
    A_{ij} = 1 \quad \textrm{ with probability } P_{c_i c_j}
\end{equation}

The above multi-community stochastic block model captures the community structure of OSNs and the variation of contact rates and recovery rates for different communities. This idea is similar to \cite{note7smilkov2014beyond} which specifies that individuals are more likely to connect with others with similar susceptibility.

In the following subsection, we construct a preferential attachment model that accounts for popular users' disproportionate role in misinformation propagation. %

\subsection{SIR Model with Preferential Attachment}
\label{subsec:stochastic PA}
Many real-world networks exhibit fat tail degree distributions \cite{jackson2010social}. In citation networks, for example, an article gains citations in proportion to the amount of citations it already has. The more citations an article has, the more likely it will be found and linked (cited) by future articles. Barabasi and Albert \cite{albert2002statisticalBAPA} denote such a link formation process as preferential attachment. We describe the SIR model with preferential attachment as follows.

Define $N_i$ as node $i$'s neighborhood and $k_i=|N_i|$ as node $i$'s degree. At each time $t$, $i$ receives $k_i$ messages\footnote{Note that the $k_i$ messages may not be unique, e.g., $i$ can view a popular user's message $j$ several times in one time step.} from its neighbors highlighting the fact that users with higher connectivity receive more messages and are more likely to be infected. For each of the $k_i$ messages, $i$ selects the sender $j$ from her neighborhood according to preferential attachment, i.e., the probability that $i$ select $j$'s message is
\begin{equation}
    \frac{k_j}{\sum_{u \in N_i} k_u}, j\in N_i
\end{equation}

Define effective infection rate\footnote{Effective infection rate is different from misinformation propagation rate (\ref{eq: propagation rate}) as it is time-invariant.} of the SIR model (\ref{eq:basic SIR}) as
\begin{equation} \label{eq: effective infection rate}
    \lambda = \frac{\beta}{\alpha}
\end{equation}

In Theorem \ref{th:preferential attachment}, we show how the SIR model with preferential attachment increases the effective infection rate. We also show how to compensate for this using SIR model parameters determined under the homogeneous mixing assumption (i.e., a completed graph).

\begin{thm}\label{th:preferential attachment}
    Denote $N$ as the number of nodes, $\mathbb{E}(k^2) = \int_{0}^{\infty} k^2 P(k) dk$ and $\mathbb{E}(k^3) = \int_{0}^{\infty} k^3 P(k) dk$ as the 2-nd and 3-rd moment of the degree distribution $P(k)$. In the SIR model with preferential attachment, the effective infection rate in a network with a degree distribution $P(k)$ has the following ratio
    \begin{equation}
    \label{eq:epidemic threshold offset}
        \frac{1}{N-1} \frac{\mathbb{E}(k^3)}{\mathbb{E}(k^2)} %
    \end{equation}
    to that in a homogeneous mixed population, i.e., a complete graph with the same number of nodes $N$.    
\end{thm}
\begin{proof}
    See Appendix \ref{appendix: theorem 3}. 
\end{proof}

The key takeaway from Theorem \ref{th:preferential attachment} is that for an SIR model with preferential attachment, the contact rate needs to multiply $(N-1) \frac{\mathbb{E}(k^2)}{\mathbb{E}(k^3)} $, to achieve the same epidemic spreading as the same model in a homogeneous population. For an SIR model without preferential attachment, in comparison, the contact rate needs to multiply $\frac{N-1}{\mathbb{E}(k)} \geq (N-1) \frac{\mathbb{E}(k^2)}{\mathbb{E}(k^3)}$. This implies that misinformation spreads to a larger network population in a preferential attachment model than an SIR model without preferential attachment.   
\vspace{0.2in}

In Section \ref{sec: numerical}, we simulate the SIR model with preferential attachment to compare the results of misinformation propagation in an OSN and a \BOSN{}.
The simulation algorithm of misinformation propagation in \BOSN{} is described in Algorithm \ref{alg:stochastic}. %
The main idea is as follows: If a user $i$ selects and contacts an infected neighbor $j$, $i$ will not become infected instantly, as in the OSN without blockchain; instead, a propagation time will be sampled from the distribution (\ref{eq:time distribution blockchain}), %
showing how long it takes for $i$ to be infected as a result of contact with $j$. The minimum value of all such propagation time is the time $i$ becomes infected. As a result, Algorithm \ref{alg:stochastic} can be implemented using a stack with $O(1)$ time complexity to return the minimum propagation time for each susceptible user. The time complexity of Algorithm \ref{alg:stochastic} is $O(N\bar{k}T)$, where $\bar{k}=\mathbb{E}(k)$ is the average degree and $T$ (Algorithm \ref{alg:stochastic}) is the number of simulation time steps.

\section{Numerical Illustration of the Blockchain Enabled Social Media Network}
\label{sec: numerical}
In this section, we simulate misinformation propagation in a three-community network using the SIR model with preferential attachment as discussed in Section \ref{sec: PA}. The simulation results show that \BOSN{s} can reduce the number of users affected by misinformation (Definition 2) and slow down the misinformation propagation rate (\ref{eq: propagation rate}). 

We also show that with the same proportion of honest miners in the blockchain system, some communities are more resilient to misinformation due to higher recovery rate and lower contact rate. This indicates that Cannikin's Law\cite{tilman2020resource} (wooden bucket theory) hold, namely, the minimum proportion of honest miners to avoid all the users getting infected is determined by the community most fragile to misinformation.

\subsection{Parameter Estimation from Twitter Datasets}
\label{subsec:parameter estimation}
In the simulation below, we use Twitter datasets of trending hashtags\cite{skaza2017modeling} to obtain realistic SIR model parameters for simulation. We justify this choice of dataset for the following reasons: 
Twitter hashtags are used to describe virally popular events and subjects. The spread of viral hashtags frequently follows a similar pattern to that of diseases\cite{weng2013virality}. 

A total of 4574 time-stamped tweets were collected during a 87-minute period using Twitter API by querying the hashtag "\#BBWLA" in 2016\cite{skaza2017modeling}. The hashtag is about an American reality television series \textit{Basketball Wives}. We chose this hashtag because %
the linked tweets provide us with an approximation of how misinformation spreads\footnote{As per the social media monitoring tool: \url{https://brand24.com/blog/what-is-media-monitoring-and-analysis/}, the number of tweets with "\#BBWLA" hashtag in the last 24 hours (May 18th, 2022) is 86. In 2016, 4574 tweets were sent in less than 2 hours, hence the tweets can be considered viral in 2016 and useful in representing misinformation spread.}. The tweets are grouped into one-minute time windows resulting in a smoothed time series of the infected population $I_{obs}(t), t=0,\cdots, 87$.

We use Bayesian Markov Chain Monte Carlo (MCMC)\cite{skaza2017modeling} to estimate the SIR parameters, $\beta$ and $\alpha$, as well as the inital values for the infected ($I(0)$) and susceptible ($S(0)$) populations. We assign the parameters with uniform priors:
\begin{equation}
\label{eq:MCMC prior}
\begin{split}
    \pi(\beta) = \textrm{U}(0, 1), \quad \pi(\alpha) = \textrm{U}(0, 1), \\
    \pi(I(0)) = \textrm{U}\left(1, \max(I_{obs}(t))\right),  \\
    \pi(S(0)) = \textrm{U}\left(\max\left(I_{obs}\left(t\right)\right), 40000 \right) \\
\end{split}
\end{equation} 
The upper bound of the initial susceptible population was chosen as 40000. For the purpose of our MCMC simulations this is adequate since the simulations do not yield posterior estimates larger than 400. We also estimate the standard deviation of the observation error $\sigma_{I_{obs}}$ (abbreviated as $\sigma_{I}$), for which we choose the Jeffreys non-informative prior\footnote{Jeffreys noninformative prior is used here because we have litte prior information about the noise in the Twitter dataset.} %
\begin{equation}
\label{eq:MCMC prior 2}
    \pi(\sigma_{I}) \propto \frac{1}{\sigma_{I}}
\end{equation}

\begin{table} 
\caption{Posterior mean and quantiles of SIR model parameters estimated from the Twitter hashtag "\#BBWLA" dataset}
\label{table:real SIR}
\begin{threeparttable}
  \centering 
  \begin{tabular}{ |C{0.094\textwidth}|C{0.094\textwidth}|C{0.094\textwidth}|C{0.094\textwidth}|} 
    \hline
    Parameters  & 0.05 Quantile & \textbf{Mean} & 0.95 Quantile  \\ %
    \hline
    $\beta$ & 0.00351 & \textbf{0.00359} &0.00369 \\ \hline
    $\alpha$ & 0.02138 &\textbf{0.02162} &0.02186 \\ \hline
    $I(0)$ & 4.17681 &\textbf{4.51518} &4.85865 \\ \hline
    $S(0)$ & 114.75589 &\textbf{115.46980} &116.27142 \\ \hline
    $\sigma_I$ & 14.74463 &\textbf{14.88995} &15.04935 \\
    \hline
    \end{tabular}
\end{threeparttable}
\end{table}

Let $\Psi=\{\beta, \alpha, I(0), S(0), \sigma_{I} \}$ be the set of SIR model's parameters. Given SIR model's difference equation
\begin{equation}
    I(t) = I(t-1) + \beta S(t-1)I(t-1) - \alpha I(t-1)
\end{equation}
we can compute the likelihood of the observed data $I_{obs}$
\begin{equation}
    L(I_{obs}|\Psi)  = \prod_{t=1}^{87} 
\frac{1}{\sigma_I\sqrt{2\pi}} 
  \exp\left( -\frac{1}{2}\left(\frac{I(t)-I_{obs}(t)}{\sigma_I}\right)^{\!2}\,\right)
\end{equation}
The posteriors of $\Psi$ is updated according to
\begin{equation}
    p(\Psi) \propto \pi(\Psi) L(I_{obs}|\Psi)
\end{equation}

\begin{figure}
     \centering
     \begin{subfigure}[b]{0.5\textwidth}
        \centering
        \includegraphics[width=1\textwidth]{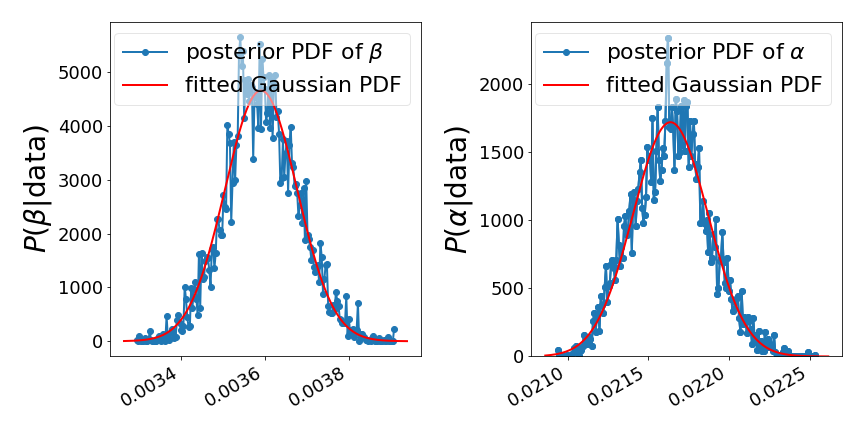}
        \caption{Posterior distribution of the MCMC samples for the contact rate $\beta$ and recovery rate $\alpha$. The mean, the 5th and 95th percentiles of the samples are displayed in Table \ref{table:real SIR}; A fitted Gaussian probability density curve with mean and standard deviation is also shown. $\beta$ is fitted by $\mathcal{N}(0.0036, 1e-4)$ and $\alpha$ is fitted by $\mathcal{N}(0.022, 0.00024)$. The posterior means are used as the SIR model parameters in the numerical simulation.} 
        \label{subfig:mcmc}
     \end{subfigure}
     \hfill
     \begin{subfigure}[b]{0.5\textwidth}
        \centering
        \includegraphics[width=1\textwidth]{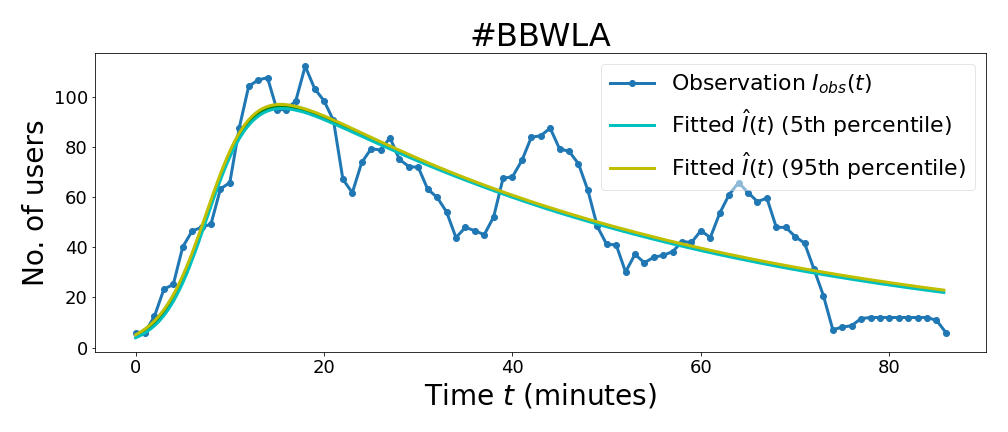}
        \caption{The theoretic value computed using the fitted SIR model and the observed infected population. The dataset is for Twitter hashtag "\#BBWLA".}
        \label{subfig:SIR real data}
    \end{subfigure}
    \caption{The SIR model (\ref{eq:basic SIR}) with parameter $\beta$ and $\alpha$ estimated using MCMC fits the Twitter dataset.}
    \label{fig:mcmc}
\end{figure}

The estimation results for the "\#BBWLA" hashtag dataset is shown in Table \ref{table:real SIR}. The MCMC simulation is run for 10000 iterations with a burn-in of 75\%. Fig. \ref{subfig:mcmc} displays the posterior sample distribution plots for the estimated $\beta$ and $\alpha$. The infected population curves computed using the estimated parameters is shown in Fig. \ref{subfig:SIR real data} along with the observation data.

\subsection{Numerical Simulation of SIR Model with Preferential Attachment in an OSN and a \BOSN{}} \label{subsec:numerical}
We use parameters estimated from Twitter dataset (Section \ref{subsec:parameter estimation}) to replicate the SIR model with preferential attachment defined in Section \ref{sec: PA}. Under the model, we investigate misinformation propagation in an OSN and a \BOSN{}, respectively. The code is publicly available at \url{https://tinyurl.com/sir-blockchain}.

\vspace{0.05in}
\noindent
{\bf Model Parameters:}
The number of simulation time steps is $T=600$. The total number of users in the simulation is $N=300$, with $100$ users in each of the three communities. The $3\times 3$ block model matrix (\ref{eq:SBM}) is
\begin{equation}
    P = 
    \begin{bmatrix}
    0.04 & 0.004 & 0.004\\  0.004 & 0.04 & 0.004\\  0.004 & 0.004 & 0.04
    \end{bmatrix}
\end{equation}
which corresponds to the situation where users have more intra-community connections than inter-community ones. 

According to Theorem \ref{th:preferential attachment} (Section \ref{subsec:stochastic PA}), we offset the contact rate by multiplying it by $(N-1) \frac{\mathbb{E}(k^2)}{\mathbb{E}(k^3)}$, where $\mathbb{E}(k^2)$ and $\mathbb{E}(k^3)$ denote the 2-nd and 3-rd moments of the degree distribution. %
In the simulated network, $\mathbb{E}(k^2) = 23.03$, $\mathbb{E}(k^3) = 139.93$. The corrected contact rate is thus $0.178$. %
We construct three communities with different contact rates 
\begin{equation}\label{eq:simulation beta}
    \beta_1 = 0.036, \beta_2 = 0.178, \beta_3 = 0.889
\end{equation}
and recovery rates 
\begin{equation}\label{eq:simulation alpha}
    \alpha_1=0.1, \alpha_2 =0.022, \alpha_3=0.005
\end{equation}
based on the parameters obtained from the Twitter dataset. We initialize the infected population by randomly choosing 3 users in each community.

Because the network maintains one unique blockchain, the three communities share the same propagation time distribution $P(T_m\leq s)$. We set the block confirmation $b$, fraction of dishonest miner $\frac{\mu_d}{\mu_d+\mu_h}$, the mining rate $\mu$, and the conditional probabilities $p_0$, $p_1$ as follows:
\begin{equation}\label{eq:simulation parameters}
    \frac{\mu_d}{\mu_d+\mu_h}=0.3, \mu = \mu_d+\mu_h = 2, p_0=0.8, p_1=0.4
\end{equation}

\begin{figure}
	\centering
	\includegraphics[width=0.48\textwidth]{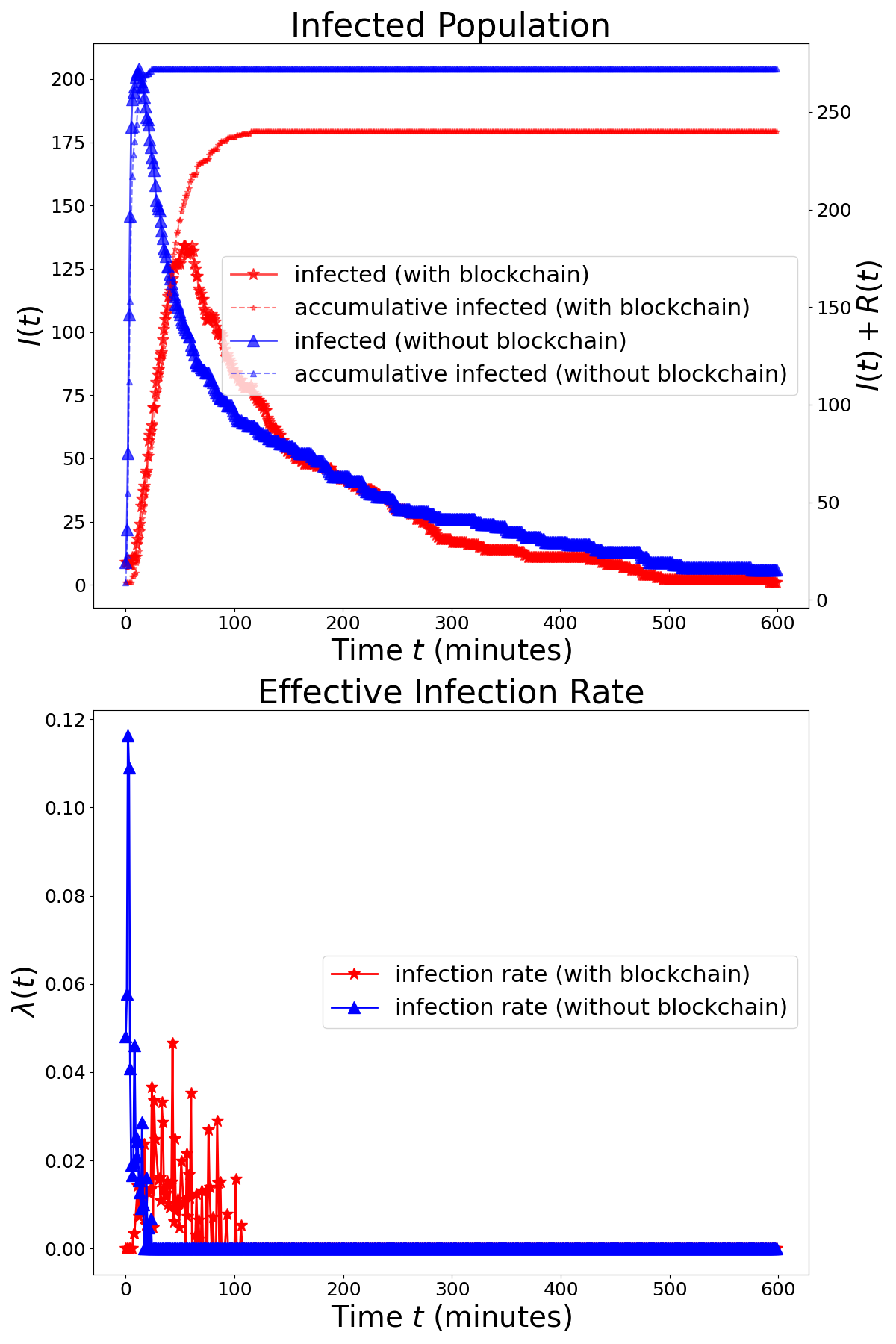}
	\caption{The simulation result of SIR model with preferential attachment in an OSN and a \BOSN{}. The top subfigure shows the infected population $I(t)$ while the bottom subfigure shows the misinformation propagation rate $\lambda(t)$ (\ref{eq: propagation rate discrete}). \BOSN{} (red star) flattens the curve of the infected population and has a smaller propagation rate compared with the network without blockchain protocol (blue up triangle). The main takeaway is that blockchain enabled network has stronger resilience to misinformation.}
	\label{fig:SIR}
\end{figure}

\vspace{0.05in}
\noindent
{\bf Performance Metrics:}
We adjust misinformation propagation rate (\ref{eq: propagation rate}) to measure the performance in mitigating misinformation in discrete time. The empirical contact and recovery rate are
\begin{equation}
    \begin{split}
        \beta(t) & = -\frac{S(t+1)-S(t)}{S(t)I(t)} \\
        \alpha(t) & = \frac{R(t+1)-R(t)}{I(t)} \\
    \end{split}
\end{equation}
We define misinformation propagation rate in discrete time as
\begin{equation} \label{eq: propagation rate discrete}
    \lambda(t) = - \frac{S(t+1) - S(t) }{S(t) (R(t+1) - R(t))}
\end{equation}

\begin{figure}
	\centering
	\includegraphics[width=0.5\textwidth]{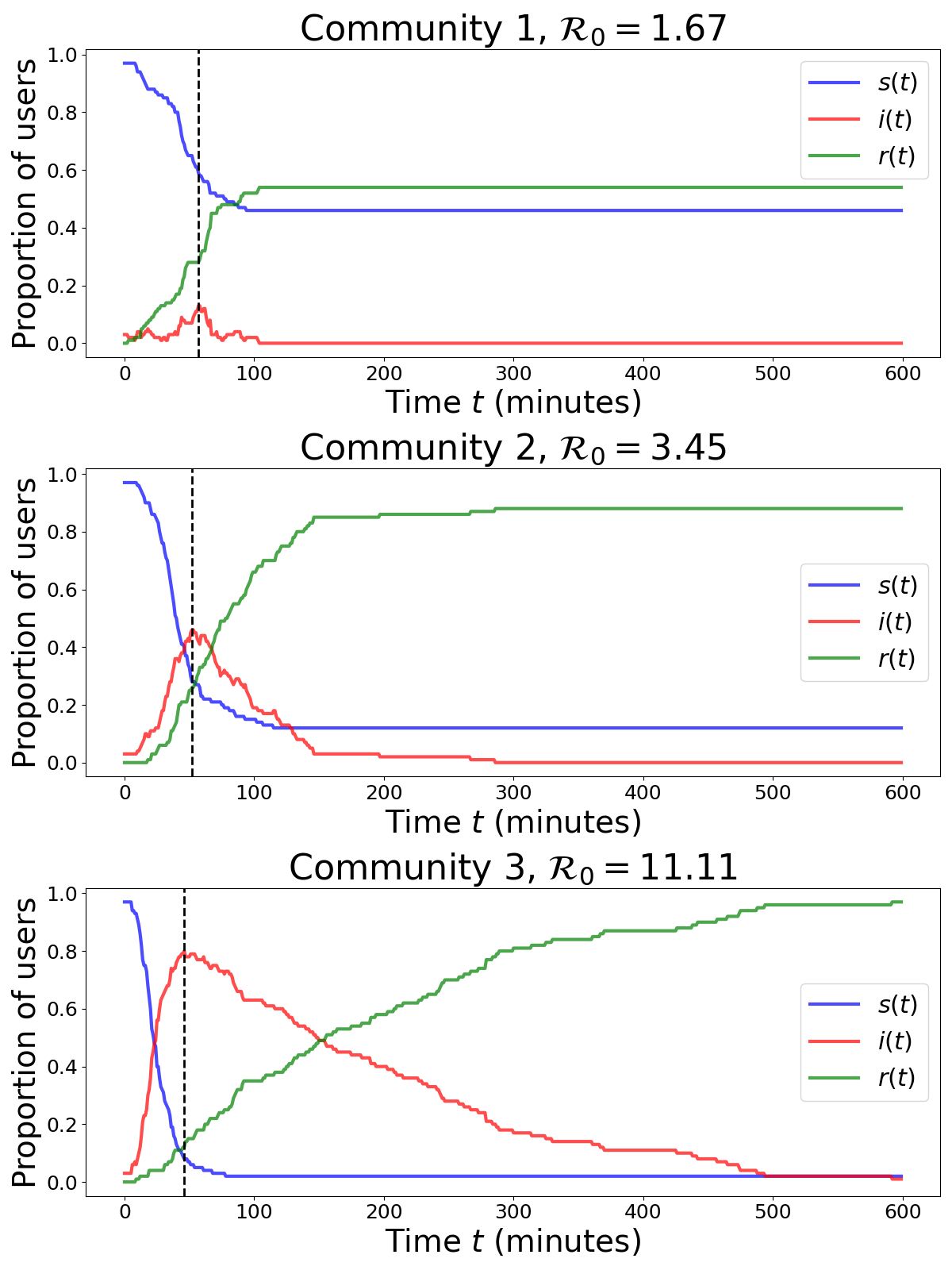}
	\caption{The simulation result of SIR model with preferential attachment in each community of a \BOSN{}. The curves represent the susceptible proportion $s(t)$ (blue), the infected proportion $i(t)$ (red), and the recovered proportion $r(t)$ (green). The black dashed vertical line denotes $t^{*}$ when the infected population reaches its peak, which is used to compute the reproduction number $\mathcal{R}_0$ defined in (\ref{eq:reproduction}). Community 1 has the smallest $\mathcal{R}_0$ while community 3 has the largest one. %
 The key takeaway from this figure is: communities will react differently to the dissemination of misinformation in a multi-community \BOSN{}. To prevent infecting all users in the community most vulnerable to misinformation, a minimal proportion of honest miners $\frac{\mu_h}{\mu_d + \mu_h}$ can be chosen accordingly.}
	\label{fig:community}
\end{figure}

In addition, to compare the performance of different communities in coping with misinformation, we compute the reproduction number $\mathcal{R}_0$ of each community. 
$\mathcal{R}_0$ is the average number of secondary cases produced by one infected individual introduced into a population of susceptible individuals\cite{van2017reproduction}. $\mathcal{R}_0 > 1$ implies that the epidemic (misinformation) can not die out without external control. The reproduction number $\mathcal{R}_0$ is computed as follows\cite{bertozzi2020challenges}:
\begin{equation}
\label{eq:reproduction}
    \mathcal{R}_0 = \frac{N}{S(t^*)},
\end{equation}
where $t^* = \argmax_{t=0,\cdots,T} I(t)$, i.e., the time when the infected population reaches the peak.

\begin{figure}
	\centering
	\includegraphics[width=0.5\textwidth]{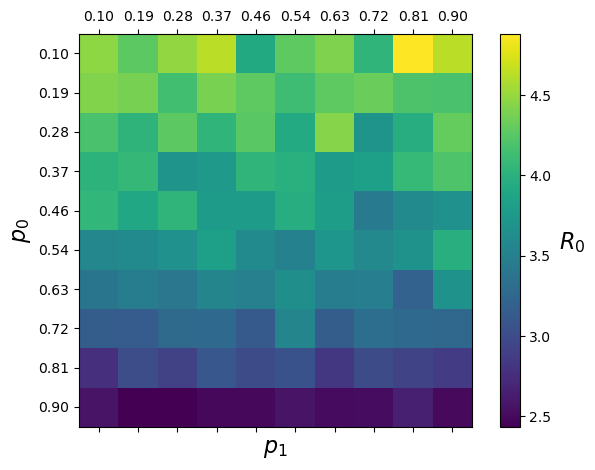}
	\caption{The reproduction number $\mathcal{R}_0$ with different conditional probabilities $p_0$ and $p_1$ (\ref{eq:cond}). Lower $p_0$ will significantly raise $\mathcal{R}_0$ and exacerbate misinformation propagation.}
	\label{fig:R0}
\end{figure}

We also evaluate the effects of the conditional probabilities $p_0, p_1$ defined in (\ref{eq:cond}) (Recall $p_0$ is the conditional probability of truthful message given a normal posting and $p_1$ is the conditional probability of truthful message given a double-spend attack) on the spread of misinformation. Figure \ref{fig:R0} shows the network's reproduction number $\mathcal{R}_0$ with different combinations of $p_0$ and $p_1$ (with other parameters fixed as in (\ref{eq:simulation parameters})). We run 10 simulations with one set of $p_0$ and $p_1$, and record the average $\mathcal{R}_0$.  Lower $p_0$ will significantly raise $\mathcal{R}_0$ and exacerbate misinformation propagation because a normal posting is more likely to be a misinformation with smaller $p_0$ (greater false negative rate). In order to reduce misinformation, the conditional probability model should be more sensitive and have a high true positive rate.

\vspace{0.05in}
\noindent
{\bf Results:}

\begin{list}{\labelitemi}{\leftmargin=0em}
\item \textit{Blockchain Improves a Network's Resilience:}  Figure \ref{fig:SIR} compares the population dynamics and the misinformation propagation rate in an OSN and a \BOSN{}. In the \BOSN{}, curve of $I(t)$ is flattened, with lower number of infected users and the infections distributed along a longer time. The misinformation propagation rate is also smaller. This suggests that the \BOSN{} slows down the misinformation propagation rate, and lowers the number of the infected users. 
It should be noted that misinformation spreads using the preferential attachment model. Regardless of the user's popularity or community membership, miners validate postings in a decentralized manner according to the Protocol \ref{alg:BOSN}. As a result,  the Matthew effect in social networks has less of an impact on the \BOSN{}.

\item \textit{Different Communities Behave Differently to Misinformation:}  Figure \ref{fig:community} compares the performance of the three communities in the \BOSN{}. Specifically, it displays the fraction of each population, with $s(t)=\frac{S(t)}{N}$ representing the fraction of susceptible individuals in the community; $i(t)$ and $r(t)$ are defined similarly. The reproduction number $\mathcal{R}_0$ is also computed for each community.

Due to the chosen parameters in (\ref{eq:simulation beta}, \ref{eq:simulation alpha}), community 1 has the strongest resilience to misinformation. The reproduction number $\mathcal{R}_0=1.67$ is the smallest among the three communities; the peak value of infected portion is only about 0.1, 0.5 of the total population is unaffected in the end, and the misinformation exist for around 100 minutes, much shorter than that of community 3. On the other hand, community 3 has the weakest resilience: the peak value of infected portion is nearly 0.8. %
The wooden bucket theory posits that the minimum fraction of honest miners in the \BOSN{} is determined by the community 3, which is most vulnerable to misinformation. %
\end{list}

\section{Conclusions}
\label{sec:conclusion}
We proposed a blockchain protocol (Protocol \ref{alg:BOSN}) to mitigate the spread of misinformation in an OSN. Our key idea was to model misinformation as a double-spend attack in blockchain. We developed a conditional probability model (\ref{eq:cond}) to account for the fact that a double-spend attack does not necessarily entail misinformation. The misinformation propagation time under the conditional probability model is distributed as a first hitting time of a biased random walk, according to Theorems \ref{th:misinformation block}.

We constructed a SIR model with preferential attachment to describe the spread of misinformation. According to Theorem \ref{th:preferential attachment}, misinformation is more likely to spread to a larger network population for a SIR model with preferential attachment than for a model without.  In numerical studies, we used Twitter hashtag datasets to estimate the SIR model parameters using a Bayesian MCMC approach. %
The result demonstrated that the proposed Protocol \ref{alg:BOSN} delays the spread of misinformation by requiring a longer confirmation time. It reduces the spread of misinformation and flattens the infected population curve.

\ifCLASSOPTIONcompsoc
  \section*{Acknowledgments}
\else
  \section*{Acknowledgment}
\fi

This research was supported by the  U. S. Army Research Office under grants W911NF-19-1-0365, U.S. Air Force Office of Scientific Research under grant FA9550-22-1-0016, and the National Science Foundation under grant CCF-2112457.
The authors would like to thank Yucheng Peng and Buddhika Nettasinghe for helpful discussions.

\ifCLASSOPTIONcaptionsoff
  \newpage
\fi

\bstctlcite{IEEEexample:BSTcontrol}
\bibliographystyle{IEEEtran}

\end{document}


\title{Supplemental Material for Mitigating Misinformation Spread on Blockchain Based Online Social Networks}
\author{Rui~Luo, Vikram~Krishnamurthy,~\IEEEmembership{Fellow,~IEEE},
        and~Erik~Blasch,~\IEEEmembership{Fellow,~IEEE}%
\IEEEcompsocitemizethanks{\IEEEcompsocthanksitem R. Luo is with the Sibley School of Mechanical and Aerospace Engineering, Cornell University, Ithaca, NY, 14850. \protect 
E-mail: rl828@cornell.edu
\IEEEcompsocthanksitem V. Krishnamurthy is with the School of Electrical and Computer Engineering, Cornell University, Ithaca, NY, 14850. \protect
E-mail: vikramk@cornell.edu 
\IEEEcompsocthanksitem E. Blasch is with Air Force Office of Scientific Research (AFOSR), Arlington, VA, 22203. \protect
E-mail: erik.blasch.1@us.af.mil
 \IEEEcompsocthanksitem This research was supported by the  U. S. Army Research Office under grants W911NF-19-1-0365, U.S. Air Force Office of Scientific Research under grant FA9550-22-1-0016, and the National Science Foundation under grant CCF-2112457. }}

\maketitle
\appendices

\begin{algorithm}
\caption{\textbf{Stochastic simulation of the SIR dynamics in the multi-community network}}\label{alg:stochastic}
\textbf{Input}: 
graph $G=(V, E)$; adjacency matrix $A$; users' community label $c_i\in\{1, \cdots, M\}, i\in V$; contact rates $\beta_m$ and recovery rates $\alpha_m, m=1,\cdots, M$; number of simulation time steps $T$.

\textbf{Output}: $S(t), I(t), R(t)$; $\mathcal{S}(t), \mathcal{I}(t), \mathcal{R}(t)$.

\begin{algorithmic}[1]
\State Three nodes are selected uniformly at random from each community to be the first infected individuals $\mathcal{I}(0)$.
\State All other nodes are initialized as susceptible with its state represented as a tuple, i.e., $\mathcal{S}(0)=\{(i, t^{(S)}_{i}) | i \in V \setminus \mathcal{I}(0) \} $, where $t^{(S)}_{i} = \infty$\footnotemark{} denotes an infinite propagation time.

\For{$t=1, \cdots, T$} %
    \For{$(i, t^{(S)}_i) \in \mathcal{S}(t-1)$} 
    \Comment{\textcolor{blue}{$S\rightarrow I$}}
        \If{$t^{(S)}_i = \infty$}
            \For{$n=1,\cdots, k_i$}
                \State Choose one of $i$'s neighbor $j$ with probability
                \[\frac{k_j}{\sum_{u: A_{iu}=1} k_u}\]
                \If{$j$ is infected}
                    \State Sample $U\sim \textrm{Unif}(0,1)$
                    \If{$U \leq \beta_{c_j}$}
                        \State Sample $t^{{(S)}^{*}}_i$ according to  (\ref{eq:time distribution blockchain}) %
                        \State Update $t^{(S)}_i = \min(t^{(S)}_i, t^{{(S)}^{*}}_i)$\footnotemark{} 
                    \EndIf
                \EndIf
            \EndFor
        \Else{ \If{$t^{(S)}_i <= 1$ }  
                \State $\mathcal{S} = \mathcal{S} \setminus \{(i, t^{(S)}_i)\}$ 
                \State $\mathcal{I} = \mathcal{I} \cup \{i\}$ 
                \Else{ $t^{(S)}_i = t^{(S)}_i - 1$}
                \EndIf}
        \EndIf
    \EndFor
    \State Simulate newly recovered node set $\mathcal{R}^{*}$ by one Bernoulli trial Ber($\alpha_{c_i}$) for each infected node $i$ %
    \Comment{\textcolor{blue}{$I\rightarrow R$}}
    \State $\mathcal{I} = \mathcal{I} \setminus \mathcal{R}^{*}$ 
    \State $\mathcal{R} = \mathcal{R} \cup \mathcal{R}^{*}$
\EndFor
\end{algorithmic}
\end{algorithm}
\addtocounter{footnote}{-2} %
\stepcounter{footnote}\footnotetext{$t^{(S)}_{i}$ denotes the time that node $i$ will remain in $\mathcal{S}$. It is initialized as $\infty$ to denote that node $i$ has not contacted with infected users.}
\stepcounter{footnote}\footnotetext{In other words, $i$ will sample one propagation time every time she contacts with an infected user, and her propagation time will be the minimum among all the samples.}

\section{Proof of Proposition \ref{prop:hitting probability}}
\label{appendix: proposition 1}
Using the methods in \cite{joriki}, we define the discrete time martingale process with $M_0=1$ and

\begin{equation}
    M_j=\begin{cases}\frac{p}{1-p}M_{j-1}&X_j=-1\;,\\\frac{1-p}{p}M_{j-1}&X_j=1\;,\\\end{cases}
\end{equation}
where $p=\frac{\mu_d}{\mu_d+\mu_h}$. Recall $\mu_d$ and $\mu_h$ are the mining rates of dishonest and honest miners respectively (\ref{eq:random walk probability}). The martingale stops when one of $\pm b$ is hit. The expected value at the stopping time is $1$; the value at $-b$ is $\left(\frac{p}{1-p}\right)^b$; and the value at  $b$ is $\left(\frac{1-p}{p}\right)^b$. Thus, with $x=P(T_b < T_{-b})$, we have

\begin{equation}
    x\left(\frac{1-p}{p}\right)^b+(1-x)\left(\frac{p}{1-p}\right)^b=1\;,
\end{equation}
with the solution
\begin{equation}
\begin{split}
    x &= \frac{1-\left(\frac{p}{1-p}\right)^b}{\left(\frac{1-p}{p}\right)^b-\left(\frac{p}{1-p}\right)^b} = \frac{1-\left(\frac{1-p}{p}\right)^b}{1-\left(\frac{1-p}{p}\right)^{2b}} = \frac{1}{1 + (\frac{\mu_h}{\mu_d})^b} \;.
\end{split}
\end{equation}

\section{Proof of Theorem \ref{th:misinformation block}}
\label{appendix: theorem 1}
Consider a random walk of $p$ plus ones and $q$ minus ones. Let $n=p+q>0$ and $x=p-q$. 
Refer to rectangular coordinates $t, x$ where the $t$-axis is horizontal and the $x$-axis is vertical. The number of paths from the origin to $(n, x)$ is $\binom{n}{x}$.  %
Let $A=(a, \alpha)$ and $B=(b, \beta)$ be integral points in the positive quadrant: $b>a\geq0$, $\alpha>0$, $\beta>0$. By reflection of $A$ on the $t$-axis is meant the point $A'=(a, -\alpha)$.

To prove Theorem \ref{th:misinformation block}, we use the Reflection principle. 

\vspace{0.2in}
\noindent
{\bf Lemma. (Reflection principle\cite{feller2008introduction})} 
The number of paths from $A$ to $B$ which touch or cross the $x$-axis equals the number of all paths from $A'$ to $B$. 
\vspace{0.2in}

\begin{figure}
	\centering
	\includegraphics[width=0.5\textwidth]{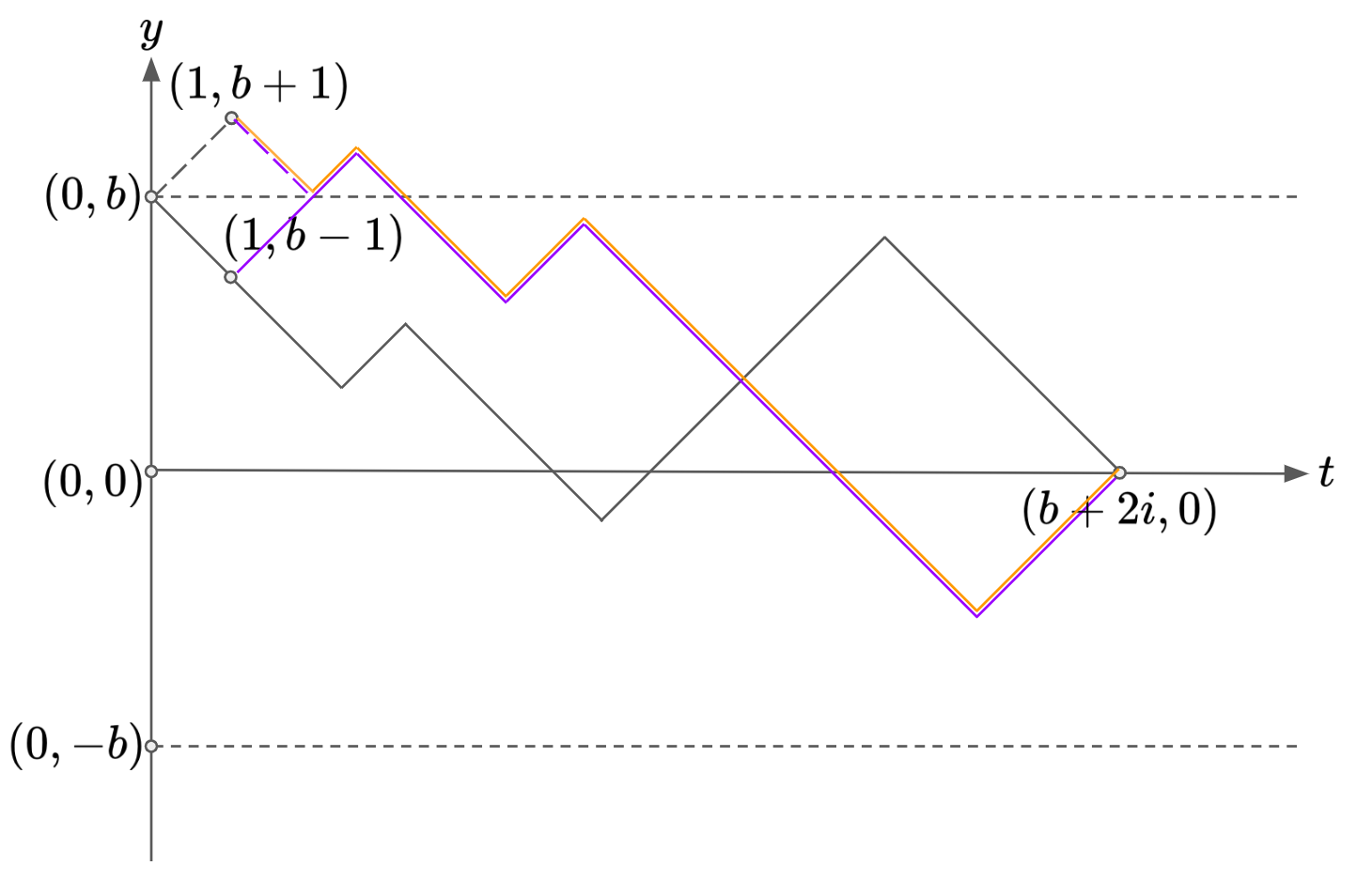}
	\caption{Consider the random walk from the origin to $(b+2i, b)$ in reverse order. By the reflection principle, the number of paths from $(1, b-1)$ to $(b+2i, 0)$ which touch or cross $y=b$, shown as a purple curve, equals the number of paths from $(1, b+1)$ to $(b+2i, 0)$, shown as an orange curve.}
	\label{fig:reflection}
\end{figure}

Now we derive the distribution of $N_b$ (\ref{eq:N_b}). $N_b=b+2i \Rightarrow S_{b+2i} = b$ and for $j<b+2i$, $-b < S_{j} < b$. 
As shown in Figure \ref{fig:reflection}, we consider $S_j$ in reverse order by flipping the paths horizontally and setting the first step as -1 (otherwise $\exists j<b+2i, \; S_j>=b$). The number of paths from $(1, b-1)$ to $(b+2i, 0)$ is $\binom{b+2i-1}{i}$.

By reflection principle, the number of paths that touch or cross $y=b$ is $\binom{b+2i-1}{i-1}$. Similarly, the number of paths that touch or cross $y=-b$ is $\binom{b+2i-1}{i-b}$. 
Now consider the paths that touch or cross both $y=b$ and $y=-b$. There are $\binom{b+2i-1}{i-b-1}$ paths that first touch or cross $y=b$ and then touch or cross $y=-b$. And the number of paths that first touch or cross $y=-b$ and then touch or cross $y=b$ is $\binom{b+2i-1}{i-2b}$. 

According to the inclusion-exclusion principle, the number of admissible paths is $\binom{b+2i-1}{i} - \binom{b+2i-1}{i-1} - \binom{b+2i-1}{i-b} + \binom{b+2i-1}{i-b-1} + \binom{b+2i-1}{i-2b}$. 
Thus the PMF of the first hitting time in case \RNum{1} is
\begin{equation}
\begin{split}
    &P(N_b=b+2i) \\
    =& P(S_{b+2i}=b) P(N_b=b+2i|S_{b+2i}=b) \\
    =&  p^{b+i} (1-p)^{i} \biggl[\binom{b+2i-1}{i} - \binom{b+2i-1}{i-1} \\
    &  - \binom{b+2i-1}{i-b} + \binom{b+2i-1}{i-b-1} + \binom{b+2i-1}{i-2b} \biggr] , i \in \mathbb{N}
\end{split}
\end{equation}

\section{Proof of Theorem \ref{th:preferential attachment}}
\label{appendix: theorem 3}
The proof is similar to \cite{pastor2001epidemic}. Let $\lambda$ (\ref{eq: effective infection rate}) denote the effective infection rate (without loss of generality, we set $\alpha=1$). Define $P(k)$ as the degree distribution, $\rho_k(t)$ as the probability that a node with degree $k$ is infected at time $t$.
The dynamical mean-field equation is therefore 
\begin{equation} \label{eq: mean field}
    \rho_k(t+1) - \rho_k(t) = -\rho_k(t) + \lambda k [1 - \rho_k(t)] \Theta(\rho(t))
\end{equation}
The left hand side in (\ref{eq: mean field}) denotes infected users recovering at a unit rate. The right hand side refers to users who have become infected through contact with infected neighbors, which is the product of the effective infection rate $\lambda$, the degree $k$, the probability that a node with degree $k$ is not infected\footnote{The proof assumes an SIS model and can be generalized to an SIR model.} $[1-\rho_k(t)]$, and the probability of contacting with an infected neighbor. The rate of the last event is influenced by preferential attachment and is
\begin{equation} \label{eq: rho expression}
   \Theta(\rho(t)) = \sum_{k_1} \frac{k_1 P(k_1|k) \rho_{k_1}(t) }{\sum_{k_2} k_2 P(k_2|k) }
\end{equation}
where $P(k_1|k)$ denotes the probability that an edge departing from a node with degree $k$ arrives at a node with degree $k_1$. In uncorrelated networks, i.e., each nodes chooses randomly the nodes it connects to, we have (See (7.3) in \cite{barabasi2013networkBook})
\begin{equation}
    P(k_1|k) = \frac{k_1 P(k_1)}{\mathbb{E}(k)}
\end{equation}
where $\mathbb{E}(k)$ is the average degree.
Therefore (\ref{eq: rho expression}) becomes
\begin{equation}
    \Theta(\rho(t)) = \frac{\sum_{k} k^2 P(k) \rho_k(t) }{\mathbb{E}(k^2)}
\end{equation}
In the stationary state, $\rho_k(t+1) = \rho_k(t)$. Rewritten (\ref{eq: mean field}) as
\begin{equation}
    \rho_k(t) = \frac{\lambda k \Theta}{1 + \lambda k \Theta}
\end{equation}
Thus $\Theta$ can be represented as a function of $\lambda$ as
\begin{equation} \label{eq: Theta}
    \Theta = \frac{\sum_k k^2P(k) \frac{\lambda k \Theta}{1+\lambda k \Theta}}{\mathbb{E}(k^2)}
\end{equation}
The self-consistency equation (\ref{eq: Theta}) allows a solution with $\Theta \neq 0$ and $\rho \neq 0$ when the l.h.s. and r.h.s. of (\ref{eq: Theta}) cross in the interval $0<\Theta\leq 1$. This leads to the inequality (see (9) in \cite{pastor2002immunization})
\begin{equation} \label{eq: inequality}
    \left.\frac{d}{d\Theta} \Bigg( \frac{\sum_k k^2 P(k) \frac{\lambda k \Theta}{1+\lambda k \Theta}}{\mathbb{E}(k^2)} \Bigg) \right\vert_{\Theta=0} \geq 1
\end{equation}
The critical value of $\lambda_{c_1}$ yielding the equality in (\ref{eq: inequality}) is given by
\begin{equation} \label{eq: critical 1}
    \frac{\sum_k k^2 P(k) \lambda_{c_1} k}{\mathbb{E}(k^2)} = 1 \Rightarrow \lambda_{c_1} = \frac{\mathbb{E}(k^2)}{\mathbb{E}(k^3)}
\end{equation}
On the other hand, the dynamic mean-field equation describing the node infection probability in a complete graph is
\begin{equation}
    \rho(t+1) - \rho(t) = - \rho(t) + \lambda (N-1) \rho(t) [1 - \rho(t)]
\end{equation}
which leads to the critical value as
\begin{equation} \label{eq: critical 2}
    \lambda_{c_2} = \frac{1}{N-1}
\end{equation}
Comparing (\ref{eq: critical 1}) with (\ref{eq: critical 2}) yields the ratio of effective infection rate
\begin{equation} \label{eq: threshold ratio}
    \frac{\lambda_{c_1}}{\lambda_{c_2}} = (N-1) \frac{\mathbb{E}(k^2)}{\mathbb{E}(k^3)}
\end{equation}
To get an equivalent result in a network with degree distribution $P(k)$ and under the preferential attachment model, the contact rate estimated based on the homogeneous mixing assumption must be multiplied by (\ref{eq: threshold ratio}), which is the ratio of the effective infection rate (\ref{eq: effective infection rate}). %

In comparison, for an SIR model without preferential attachment, the ratio of effective infection rate is \begin{equation}
    \frac{N-1}{\mathbb{E}(k)}
\end{equation}
which is larger than (\ref{eq: threshold ratio}) due to the moment inequality $\mathbb{E}(k^3) \geq \mathbb{E}(k) \mathbb{E}(k^2)$, indicating that misinformation spreads to a larger network population with preferential attachment.

\section{List of Blockchain Enabled Social Media Networks}
\label{appendix:list}
Below, we list several recent examples of \BOSN{s}.
\begin{list}{\labelitemi}{\leftmargin=0em}
\item Steemit (\url{https://steemit.com/}) is built on the Steem blockchain, a decentralized reward platform for publishers to monetize content and grow community. Those who hold more Steem tokens have more decision power on community matters and reward distributions.   
\item Sola (\url{https://sola.ai/}) is a hybrid of media and social network which uses AI algorithms to feed quality content to the most interested users.
\item Civil (\url{https://civil.co/}) is a community-owned network of journalists who use blockchain to establish transparency and trust. On Civil’s network, independent journalists create newsrooms where they add and share their content. 
\item onG.social (\url{https://ong.social/}) is a blockchain based social dashboard which supports community building and social interaction with cryptocurrency rewards. It runs on two blockchains, Ethereum and Wavesplatform.
\item Sapien (\url{https://www.sapien.network/}) is a social news platform built on the Ethereum blockchain that gives users control of data and content. 
\end{list}

\bstctlcite{IEEEexample:BSTcontrol}
\bibliographystyle{IEEEtran}